# Disentangling Gold Open Access

Daniel Torres-Salinas*, Nicolas Robinson-Garcia** and Henk F. Moed***

\* Universidad de Granada (EC3metrics SL y Medialab-UGR) and Universidad de Navarra, Spain
\*\* School of Public Policy, Georgia Institute of Technology, United States
\*\*\* Visiting scholar, Universidad de Granada, Spain

## Abstract

This chapter focuses on the analysis of current publication trends in gold Open Access (OA). The purpose of the chapter is to develop a full understanding on country patterns, OA journals characteristics and citation differences between gold OA and non-gold OA publications. For this, we will first review current literature regarding Open Access and its relation with its so-called citation advantage. Starting with a chronological perspective we will describe its development, how different countries are promoting OA publishing, and its effects on the journal publishing industry. We will deepen the analysis by investigating the research output produced by different units of analysis. First, we will focus on the production of countries with a special emphasis on citation and disciplinary differences. A point of interest will be identification of national idiosyncrasies and the relation between OA publication and research of local interest. This will lead to our second unit of analysis, OA journals indexed in Web of Science. Here we will deepen on journals characteristics and publisher types to clearly identify factors which may affect citation differences between OA and traditional journals which may not necessarily be derived from the OA factor. Gold OA publishing is being encouraged in many countries as opposed to Green OA. This chapter aims at fully understanding how it affects researchers' publication patterns and whether it ensures an alleged citation advantage as opposed to non-gold OA publications.

## 1. INTRODUCTION

Almost 30 years have gone by since the emergence of ArXiV, the revolutionary open access repository launched in a pre-Internet era in the early 1990s (Ginsparg, 1997). Such event indicates the first landmark on the Open Access (OA) movement, a revolution on the way scholarly works are disseminated which would not have been possible without the technological advancements that preceded them. Its spread has always been surrounded by controversy on its motivations and its effects and benefits for those providing open access. Among others, Kurtz & Brody (2006) have pointed out the *increasing access* to scientific information, identifying OA as a natural step in these ever more rapid communication processes. Contrarily, Beall (2012) sees OA publishing as a threat to the scholarly communication system, and argues that 'authors, rather than libraries, are the customers of open access publishers, so a powerful incentive to maintain quality has been removed'.

The timely expansion of OA can be explained partially by the serial breakdown (Esposito, 2004), a scholarly publishing crisis derived from the shift to online scientific publishing and the concentration of journals among a few publishers who imposed libraries to subscribe to fixed collections of journals at unsustainable prices through *big deals* (Friend, 2003); limiting access to scientific literature. However, OA publishing is now handled by few publishers who negotiate country-wide licenses for article processing charges (APC) through deals which share many similarities with the *big deal* subscription access agreements (Björk, 2016). PLOS One, Scientific Reports and Nature Communications, the three most prolific mega-journals, represent already 62.2% of all gold OA publications between 2012-2016 (Data extracted from Clarivate Analytics InCites). In the same vein, PLOS itself, generates more than $40 million annually (https://www.plos.org/financial-overview). It is safe to state therefore, that OA is no longer just an ideological movement, but also a multimillion business.

The OA movement took shape in the Budapest Open Access Initiative (http://www.budapestopenaccessinitiative.org/) in 2002 seeking to unite '[a]n old tradition and a new technology [which had] converged to make possible an unprecedented public good […]. [T]he willingness of scientists and scholars to publish the fruits of their research in scholarly journals without payment, for the sake of inquiry and knowledge'. Still, traditional publishers have constantly resisted to modify their subscription-based business model, leading to boycotts from scientists (Whitfield, 2012) and calls to defy publishers' copyright privileges (Swartz, 2008). While some concessions have been made and most journals allow nowadays preprint versions of papers to be uploaded in repositories, such defiance to break paywalls still remains, either through legal (Chawla, 2017) or illegal means (Bohannon, 2016).

Many studies and reviews have been devoted to defining its characteristics and implications (Suber, 2003, 2005; Zuccala, 2009), the state of OA (Björk, 2014; Björk et al., 2010; Harnad et al., 2008; Laakso et al., 2011), how much those OA literature represent (Archambault et al., 2014), benefits derived from it (Tennant et al., 2016; Van Noorden, 2013b) or its relation with citation impact (Eysenbach, 2006; Gargouri et al., 2010; Harnad & Brody, 2004; Moed, 2007) among others. Still, there are many questions and misunderstandings that need to be addressed to better comprehend and assess the mechanisms that are being put in place to make OA possible.

In this chapter, we focus on Gold OA, that is publications from OA journals, and analyze differences in production and impact between countries by using normalized citation scores to better characterize the phenomenon of OA publishing. Long gone is the debate questioning the viability of an OA publishing model (Crawford, 2002). Many OA journals are now renown and well-established journals and authors have accepted the APC model and are willing to pay journal publication fees. However, there are important sectors of the scientific community who still raise concerns as to the quality of these journals and the potential threats they pose to the scientific communication system (Agrawal, 2014; Beall, 2013; Bohannon, 2013).

## 2. WHAT IS OPEN ACCESS?

The Budapest Open Access Initiative (BOAI) was the one to coin the term *open access* (Tennant et al., 2016). In their founding document the offered the following definition of OA:

> *"[Research literature which is] free[ly] availab[le] on the public internet, permitting any users to read, download, copy, distribute, print, search, or link to the full texts of these articles, crawl them for indexing, pass them as data to software, or use them for any other lawful purpose, without financial, legal, or technical barriers other than those inseparable from gaining access to the internet itself. The only constraint on reproduction and distribution and the only role for copyright in this domain should be to give authors control over the integrity of their work and the right to be properly acknowledged and cited."*

The Berlin Declaration on Open Access to Knowledge in the Sciences and Humanities (BDOA) of 2003 (https://openaccess.mpg.de/Berlin-Declaration) added further specifications to this definition. First, it defined what is understood as research literature as 'original scientific research results, raw data and metadata, source materials, digital representations of pictorial and graphical materials and scholarly multimedia material'. Second, it established that OA contributions must comply with two conditions: 1) the concession of copyrights to access, use, distribute and modify freely and worldwide such documents as long the author(s) is acknowledged, and 2) that the document is deposited in an online repository which complies with certain technical standards.

But the implementation of OA led to situations which were not originally contemplated by these definitions. One of them has to do with the appearance of multiple versions of the same document (van Leeuwen, Tatum, & Wouters, 2015). As a document can be uploaded to a repository and also submitted to a journal, various versions of such paper can be available at the same time (pre-submitted manuscript, peer-reviewed manuscript and journal formatted version). This situation can create a level of uncertainty, as there could be substantial differences between these different versions. It must be noted that in many cases the version uploaded to a repository is a so-called author copy of the manuscript published in the journal, i.e., the version accepted for publication by the journal, but not subjected to the copy-editing process conducted by the publisher. This practice ensures that the scientific contents of the preprint and published version are identical.

Also, the use of relaxed notions of OA influence the perception researchers have as to what OA is and what is not. There is a tendency to consider OA and free access as synonyms. OA was formally defined more than ten years after the phenomenon had already taken place. Issues such as copyrights or how to make documents accessible were not even considered at this early stage (Crawford, 2002). It is plausible to speculate that the retroactive definition of OA has helped on this misconception of OA. For instance, a recent report commissioned by the European Commission claimed that more than half of the publications were in OA, but defined OA as 'freely available online to all (no money had to be paid, no registration to a service or website had to be made)' (Archambault et al., 2014). This was later noted by a news story published in Nature which had to emphasize that 'although free to read, [articles] may not meet formal definitions of open access because, for example, they do not include details on whether readers can freely reuse the material' (Van Noorden, 2013a). Although free access can still be perceived as positive, the fact that authors are uploading their publications to private corporations such as ResearchGate, Mendeley, figshare or Academia instead of OA repositories, raises concerns as to the future sustainability of the OA movement (Björk, 2016).

Originally, two routes to OA were envisioned to provide a middle ground that offers room to new business models while promoting sustainable and universal access to scientific literature. These are known as the green and golden routes (Harnad et al., 2008): two non-mutually exclusive models to reach OA while conceding space to journals to make profit. The green route devises scientists as the ones to be held responsible on permitting OA to their publications. They are expected to upload their works to repositories maintained by university libraries following the model set by Ginsparg and ArXiV (Björk, 2014; Cullen & Chawner, 2011). In principle, this solution leaves journals as an accessory element which ceases to be at the core of the scientific communication system. Still, journals influence authors' decision on making their work accessible as they hold the copyrights of the manuscripts they publish (Tennant et al., 2016).

The golden route maintains journals as the core of academic communication. Journals are the ones which should provide OA access. This means abandoning the subscription-based business model. Laakso et al. (2011) describe three types of OA journals: direct OA, delayed OA or hybrid OA. Direct OA journals are those which offer their full contents in OA. While most of these publishers adopt an author-pays model, including article processing charges (APC), this is not a prerequisite for direct OA journals and in fact, many journals are maintained by public institutions without incorporating any fee for authors. Delayed OA journals on the other hand, maintain a subscription-based model but compromise to offer OA access to all their contents after a given period. Hybrid OA journals are the most restrictive of the three types. In fact, many argue that they are not true OA journals as they will only provide OA access to those publications for which the authors have paid an OA clause, the rest of their contents remaining behind paywalls.

Since these two routes were defined, other types of OA have been described in the literature. We must note that some of these denominations are controversial and even not considered by some authors as truly OA, as they do not fit to the general definition provided by the Budapest Open Access Initiative. For instance, Suber (2008) makes the distinction between Gratis and Libre OA. He defines the former as that which only offers rights to *read* articles, whereas the latter extends rights to reuse articles. Another proposed type of OA is that named as Black OA (Björk, 2017), which is defined as that where articles are illegally distributed through pirate sites such as Sci-Hub and LibGen, something considered as not real OA by many. Finally, it is worth discussing Bronze OA (Piwowar et al., 2017). While not truly OA, articles falling under this category are those which are freely distributed by publishers through their website, without including any type of Open Access license. Articles falling under this category respond to forms of delayed OA (that is, journals making accessible their archive), from journals not listed in the *Directory for Open Access Journals* (DOAJ), or articles which are temporarily offered freely by journals to promote their contents.

One of the main concerns of OA advocates has been to learn how much of the scientific literature is already in open access and how is this number growing in respect with the overall growth of scientific literature. Björk et al. (2010) established that 20.4% (up to 24% if articles in hybrid journals are included) of the literature was already OA in 2008, being 8.5% gold OA. Gargouri, Lariviere, Gingras, Carr, & Harnad (2012) found similar figures when analyzing its growth and differences by discipline. They reported that an average of 24% of the literature was in OA between 2005 and 2010 and estimated. Contrarily to Björk and his team, they found out that only 2% of the OA literature was gold OA, with Biomedical Research being the field with the highest share of gold OA (8%). A more recent study (Piwowar et al., 2017) reported that up to 28% of the literature was accessible via OA, being  Bronze OA the most common form of OA (16.2%) followed by Green OA (4.8%), Hybrid OA (3.6%) and, lastly, Gold OA (3.2%).

First studies analyzing publication and citation trends in OA tended to focus on green OA (Gargouri et al., 2010; Harnad & Brody, 2004; Moed, 2007) as in many cases, the authors of these studies were themselves advocating for green OA (Gumpenberger, Ovalle-Perandones, & Gorraiz, 2013). But journals' embargo requirements and OA releasing delays have pushed others to considered gold OA as a more promising venue (Jubb, Cook, Hulls, Jones, & Ware, 2011), leading to new studies focused specifically on OA journals. Since the beginning of the 2000s the number of OA journals grown exponentially, going from less than 750 journals in 2000 to more than 6500 in 2011 (Laakso & Björk, 2012). But the types of journals have also diversified and we can now differentiate between OA journals with or without an APC model, born vs. converted OA journals or between small and mega journals.

## 3. DISENTANGLING GOLD OPEN ACCESS

One of the first issues studies on OA journals encountered, was the difficulty to find a comprehensive list of all OA journals (Gumpenberger et al., 2013; van Leeuwen et al., 2015). Recently, both Scopus and Web of Science, have included the possibility to identify OA journals within their databases. Data and figures displayed in this chapter are retrieved from Web of Science InCites and hence some discrepancies could be found with OA journal lists provided elsewhere. According to Clarivate, OA journals in InCites are based on the DOAJ open access status[1]This must be considered when interpreting the findings displayed.

The success of OA publishing is intimately related to the expansion of digital distribution and the consequent abandonment of printed materials. According to Björk & Solomon (2012) we can distinguish four waves which took place almost simultaneously in the latter half of the 1990s and the beginning of the new century. The first one is characterized by the launch of new OA journals by individual scientists, exemplified by the Journal of Medical Internet Research, currently a world leading journal in its field. The second wave consists on the adoption of an OA model by established subscription journals. Here a pioneering journal was BMJ, starting to offer OA since 2000. A more ambitious initiative in the sense of the magnitude of journals converting to OA, is the launch of Scielo in 1997, a South American portal publicly subsidized, which gives OA access to hundreds of journals at no costs to publishers.

A game changer in OA publishing was the launch of BioMed Central (BMC, currently owned by Springer) and PLoS in 1998 and 2000 respectively. These two OA born journals are the first ones to adopt an APC model, envisioned by BMC founder, Vitek Tracz. The fourth and last wave is that representing the reaction of established publishing firms to OA proposing the hybrid model. That is, subscription-based publishers implementing an OA option at the level of individual articles, such as Springer's Open Choice, thus making journals hybrid in terms of the financing model.

The APC model is seen by many as the most sustainable solution to maintain journals as a profitable business while ensuring universal access to scientific literature. Supranational and national organizations such as the European Union, the US National Institutes of Health or the British Wellcome Trust include OA policy mandates ensuring that all publicly-funded research findings are preserved and publicly accessible. In most cases, they promote both the gold and green routes, considering that the APC model allows 'savings of up to 30%' (Cochran, 2014) compared with journal subscriptions. While it can be claimed that gold OA ensures the publication of peer-reviewed papers and hence the credibility of research, others have questioned if an author-pays model really ensures the quality of the work published (Bohannon, 2013). As authors become essential to the business model of these journals, quality levels may drop or even be non-existent, leaving room for *predatory journals* which will publish anything as long as profits keep raising (Beall, 2012). This phenomenon, which is not directly related with gold OA but with the APC model, has made many researchers to perceive Gold OA publications as research of a lesser quality (Agrawal, 2014).

In the late 2000s a debate emerged in the literature with regard to a perceived OA citation advantage (Brody, Harnad, & Carr, 2006; Craig, Plume, McVeigh, Pringle, & Amin, 2007; Gargouri et al., 2010; Harnad & Brody, 2004; Moed, 2007). Green OA advocates promoted the idea that researchers who made their publications OA had a citation advantage as opposed to those who did not due to the higher accessibility and visibility of their work. While such perception was confirmed by most studies (Swan, 2010), some authors pointed out that this perception could be more of an acceleration on the number of citations rather than an advantage due to an *early view* bias that favored OA papers (Moed, 2007). In the case of Gold OA the story is rather the opposite: 'open access has multiplied that underclass of journals, and the number of papers they publish' (Bohannon, 2013). The pernicious effect of *predatory publishers* and the low barriers

---

set by many OA peer-reviewed journals have led opponents to OA to state that '[t]he open-access movement has been a blessing to anyone who has unscientific ideas and wants to get these ideas into print' (Beall, 2013).

Björk & Solomon (2012) argue that OA journals are not of a lesser quality than traditional ones but are younger. In fact, OA journals in fields such as Biomedicine, or those with an APC model indexed in Web of Science have similar citation rates than subscription-based journals (Solomon, Laakso, & Björk, 2013). This is also corroborated by Gumpenberger et al. (2013), who indicate that there is an 'overall positive impact trend for Top Gold Open Access journals'.

## 3.1. Gold OA output and impact of countries and scientific fields

In this section, we will examine such arguments by analyzing the most recent trends in OA publishing for the period 2007-2016. Contrarily to the study by Björk & Solomon (2012), we do not distinguish between OA funded by article processing charges (APC) and other, mainly subsidized forms of OA. In addition, rather than categorizing journals according to the *country of the publisher*, and calculating journal impact factor-like citation rates for a journal as a whole, the current study also presents analyses on the *article production of countries in OA journals*, as expressed in the affiliations of publishing authors, and so called *category normalized citation rates*, comparing the citation impact of an entity – e.g., the total collection of article published in OA papers, or the OA articles published by authors from a particular country – to the world citation average in the subject field in which the entity is active.

Table 1. Overview of gold Open Access publications: Evolution and distribution by research fields. 2007-2016 period

**1A. Evolution of Gold Open Access**

| Publication year | # Articles Web of Science | # OA Articles Web of Science | % OA Articles Web of Science | Category Normalized Citation Impact |
|---|---|---|---|---|
| 2007 | 1071782 | 34752 | 3,24% | 0.81 |
| 2008 | 1158136 | 50616 | 4,37% | 0.76 |
| 2009 | 1228957 | 61143 | 4,98% | 0.79 |
| 2010 | 1284685 | 73543 | 5,72% | 0.81 |
| 2011 | 1373833 | 93273 | 6,79% | 0.84 |
| **Total 2007-2011** | **5045611** | **278575** | **5,52%** | **0.81** |
| 2012 | 1414483 | 114252 | 8,08% | 0.86 |
| 2013 | 1484570 | 143753 | 9,68% | 0.89 |
| 2014 | 1527771 | 166465 | 10,90% | 0.88 |
| 2015 | 1555307 | 180337 | 11,59% | 0.83 |
| 2016 | 1466589 | 177251 | 12,09% | 0.73 |
| **Total 2012-2016** | **7448720** | **782058** | **10,50%** | **0.83** |

**1B. OECD Research Fields Gold Open Access**

| OECD Research Field | Articles Web of Science | OA Articles Web of Science | % OA Articles Web of Science | Category Normalized Citation Impact |
|---|---|---|---|---|
| NATURAL SCIENCES | 3579626 | 367004 | 10,25% | 0.97 |
| ENGINEERING | 1820952 | 123131 | 6,76% | 0.60 |
| HEALTH SCIENCES | 2238476 | 319198 | 14,26% | 0.79 |
| AGRICULTURAL SCIENCES | 353675 | 39092 | 11,05% | 0.46 |
| SOCIAL SCIENCES | 835437 | 34917 | 4,18% | 0.84 |
| HUMANITIES | 347368 | 8660 | 2,49% | 0.54 |

**Technical Note** Dataset: InCites Dataset; Schema: OECD; Document Type: Article; Time Period: 2007-2016
*InCites dataset updated May 13, 2017. Includes Web of Science™ content indexed through Mar 31, 2017.*

Table 1A provides insight into the global development of Gold OA publishing. The table shows a steady increase of this percentage over the years, from 3.24% in 2007 to 12.09% in 2016. For the period 2012-2016 this percentage amounts to 10.5%. The last column gives the category normalized citation impact of the OA articles published in the various years, correcting for differences not only in citation practices between subject fields, but also between document types and publication years. A value of 1.0 means that Gold OA articles are cited on average as frequent as an average article (either Gold OA or non-Gold OA). The category normalized citation impact of Gold OA articles ranges over the years between 0.73 and 0.89. There is no clear trend in the data. Aggregating data into two five-year periods, the scores in the two periods are statistically similar: 0.81 versus 0.83. In short, Gold OA articles are cited some 15% less often than an average article (either Gold OA or non-Gold OA article).

Table 1B presents a breakdown of the Gold OA articles their impact by scientific field. 41.1% of OA articles are published in journals assigned to the discipline Natural Sciences. Relative to the total number of articles in this subject field, the share of Gold OA articles is 10.25%, which is near the overall average of 10.5 indicated in Table 1. This is also the field with the largest normalized citation score (0.97), which is somewhat higher than the value of 0.83 obtained for the total set of all Gold OA articles in the database from all disciplines.

Figure 1. Total number and share of Gold Open Access publications by Web of Science subject category. 2012-2016 period

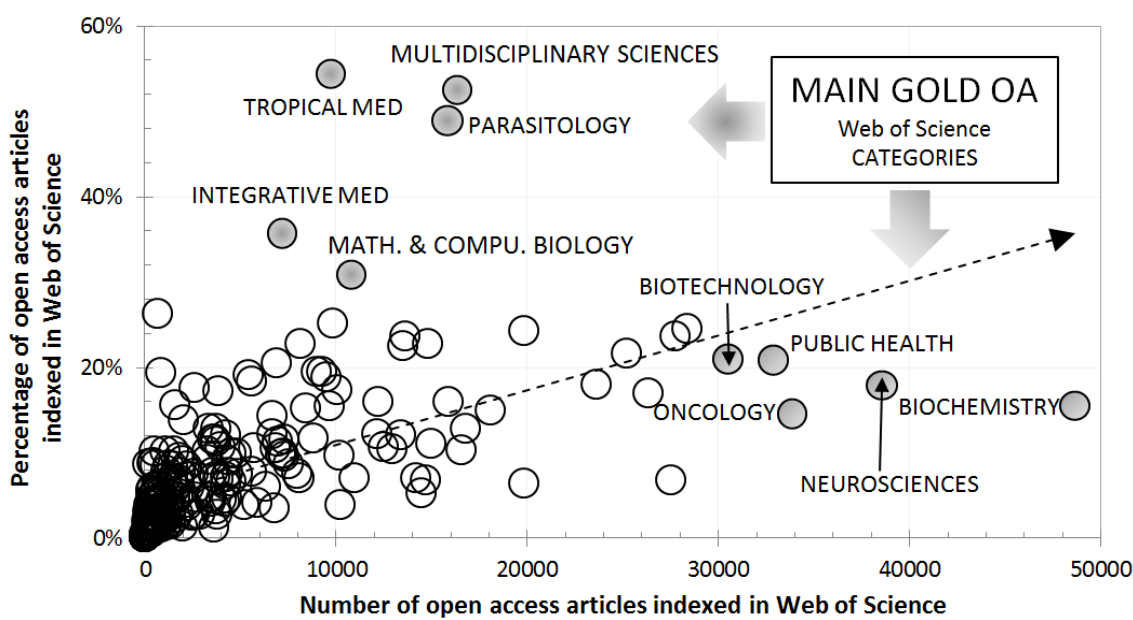

**Technical Note** Dataset: InCites Dataset; Schema: OECD; Document Type: Article; Time Period: 2007-2016
*InCites dataset updated May 13, 2017. Includes Web of Science™ content indexed through Mar 31, 2017.*

Figure 1 delves into these areas by showing the total number of Gold OA publications by Web of Science subject category and the percentage Gold OA represents within each subject category. Biochemistry is the field with the highest number of Gold OA publications, followed by neurosciences, oncology and public health. However, it is in the subject categories of multidisciplinary sciences, tropical medicine and parasitology where Gold OA represent half of the overall number of publications. The explanation of such a large share in the multidiciplinary fields is due the presence of OA mega-journals such as Plos One or Scientific Reports. In the case of Tropical Medicine and Parasitology, journals from big OA publishers are still present (e.g., Plos Neglected Tropical Diseases, Plos Pathogens or Malaria Journal, which belongs to BMC), but there are also national journals which do not publish in English language. Interestingly, these are mostly Brazilian journals like Memorias Do Instituto Oswaldo Cruz, Revista da Sociedade Brasileira de Medicina Tropical or Revista Brasileira de Parasitologia Veterinaria. This is reasonable when considering that these subject categories have a heavy local component and South America and Brazil in particular, are regions where these subject categories are more relevant in comparison with regions like North America or Europe.

The ranking of countries according to their number of Gold OA publications differs substantially from that based on total publication output (Figure 2). Some top countries in terms of Gold OA make a much smaller contribution to the total global output. For instance, Brazil, which occupies the 14th position of most productive countries within the 2007-2016 period according to the Web of Science, is actually the fourth country in terms of the absolute number of OA publications. Similarly, Spain, the 9th country with the largest number of publications within the same period, goes up to the 7th position when one considers only Gold OA publications. Another example is Mexico which falls from the 23rd position according to their contribution to OA publishing, to the 31st position when looking at their global figures.

Figure 2. Status of gold OA publications by country. 2012-2016 period. Data retrieved from Web of Science

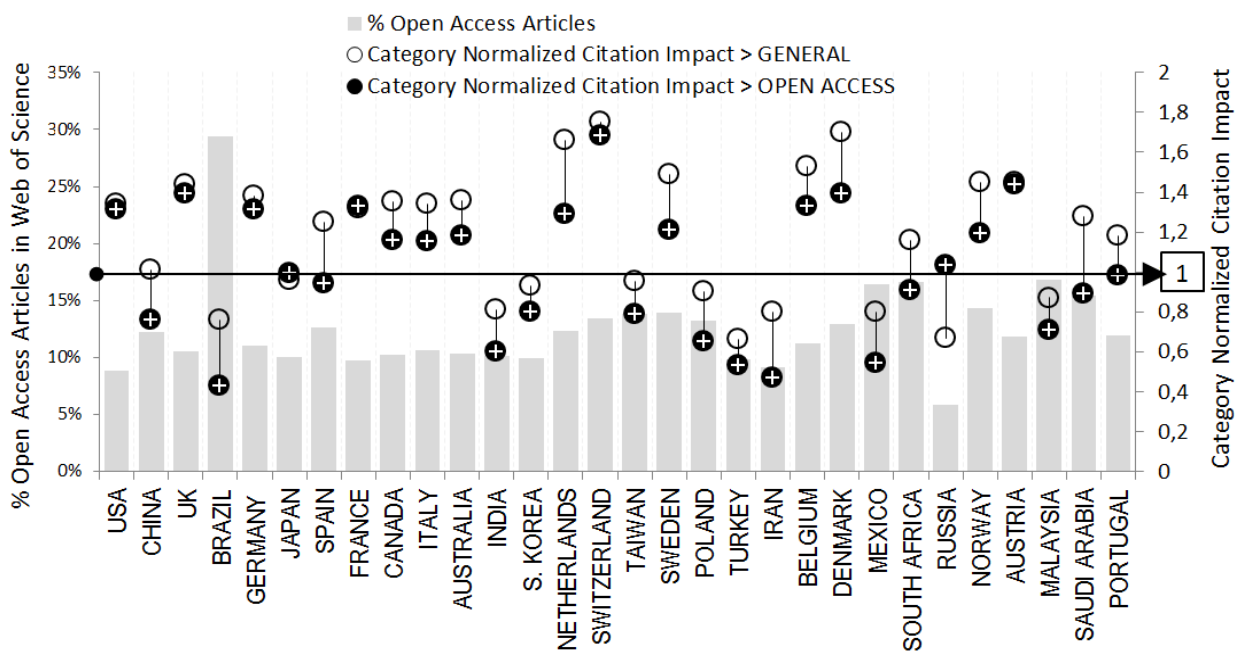

**Note and additional data:** Countries in the graph are order from left to right according the total number of Open Access articles indexed in the Web of Science: **USA** (172387), **CHN** (145691), **UK** (59921), **BRA** (58870), **GER** (56850), **JPN** (38302), **ESP** (35008), FRA (34372), **CAN** (32953), ITA (32619), **AUS** (29285), **IND** (28447), KOR (26416), **NED** (23152), **SUI** (18721), TPE (18511), **SWE** (17658), **POL** (16392), **TUR** (13400), **IRI** (12538), **BEL** (11669), **DEN** (10772), **MEX** (10405), RSA (9766), **RUS** (9061), NOR (8967), **AUT** (8580), **MAS** (8513), **KSA** (84480) and **POR** /7898). A country's percentage of OA is calculated relative to the total of articles published by that country. **Technical Note** Dataset: InCites Dataset; Schema: OECD; Document Type: Article; Time Period: 2007-2016 *InCites dataset updated May 13, 2017. Includes Web of Science™ content indexed through Mar 31, 2017.*

The case of Brazil is particularly interesting as it is by far the country with the largest share of OA publications from its overall output (almost 30%), a consequence of the gold OA proactive policy undertaken by the Brazilian government through the promotion of the SciELO platform, an initiative followed by other Latin American and Caribbean countries which provides OA access to journals from these countries (Meneghini, Mugnaini, & Packer, 2006). In the case of Spain, the share of OA publications based on its overall output is not as large as that of Brazil but, as noted elsewhere (Torres-Salinas, Robinson-Garcia, & Aguillo, 2016), Spain has increased its share of gold OA output at a higher rate than the world average over the last decade.

While for all countries the category normalized impact of OA publications is lower than the overall value of their normalized impact, the gap between these two figures differs greatly by country. Differences in the case of the United States, United Kingdom, Japan, Switzerland or Austria are almost non-existent. This is not the case for China, Brazil, Spain, France or Canada, for which the impact of Gold OA publications is substantially lower than their overall impact level. This difference can be partly explained by the disciplinary profile of these countries.

Figure 3 shows the five most productive countries of Gold OA publications by the six OECD scientific fields. Colors represent the value of the Category Normalized Citation Impact of each country-field combination. As observed, countries with higher normalized impact tend to publish most of their outputs in the medical and health sciences, the natural sciences, and social sciences. In these fields, most of the countries in the top five positions reach an overall impact equal or above of the world average (Category Normalized Citation Impact ≥ 1.0).

This is not the case in agricultural sciences, engineering and technology and humanities. All countries -  except the USA in engineering and technology and in the humanities, have lower values of normalized impact than the world average. Also, these are the only fields where the United States is not the largest contributor. Brazil, for instance, is by far the country with the largest number of publications in agricultural sciences. A closer examination in this field shows that six of the top ten journals producing most of the publications are Brazilian, all of which exhibit low Impact Factor values under 0.6. China is the number one in the case of engineering and technology, and Spain in the case of the humanities. It is also worth noticing the asymmetry of the share of publications in figures 3A and 3B, where there is a large difference between the number of publication of the first country with respect to the rest.

Figure 3. Top five most productive countries in Open Access journals by field

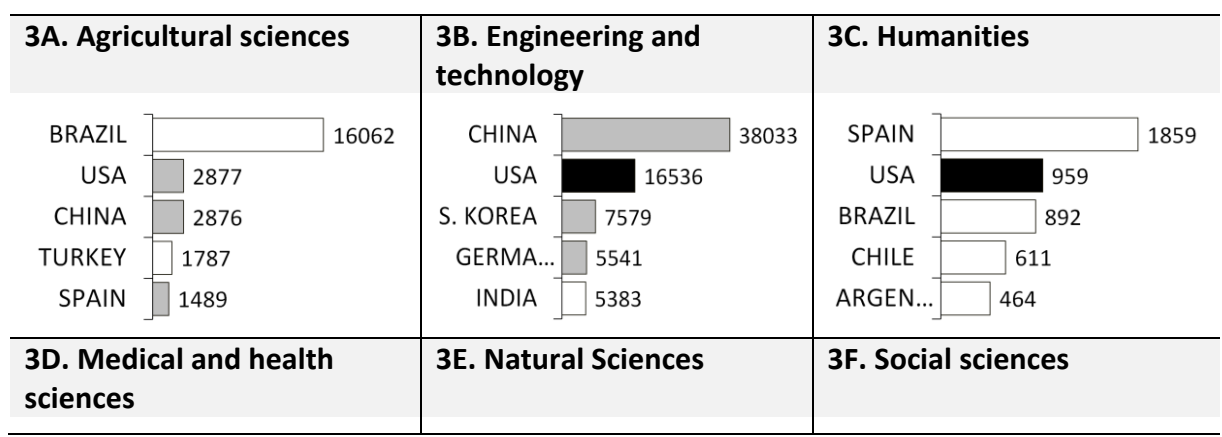

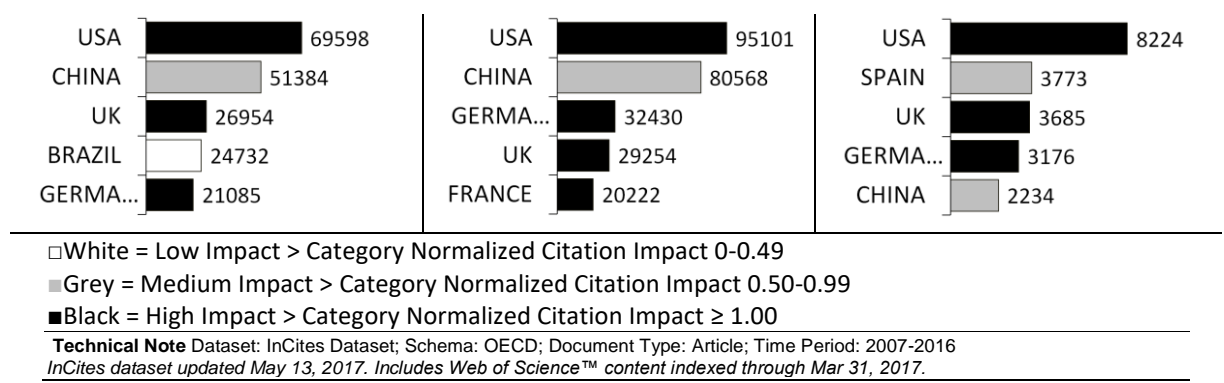

□White = Low Impact > Category Normalized Citation Impact 0-0.49
■Grey = Medium Impact > Category Normalized Citation Impact 0.50-0.99
■Black = High Impact > Category Normalized Citation Impact ≥ 1.00

**Technical Note** Dataset: InCites Dataset; Schema: OECD; Document Type: Article; Time Period: 2007-2016
*InCites dataset updated May 13, 2017. Includes Web of Science™ content indexed through Mar 31, 2017.*

These figures suggest that the differences in impact between Gold OA papers and a country's total publication output shown in Figure 2 are largely due to the Gold OA disciplinary profile of each country and the characteristics of OA journals in these different fields. This idea is reinforced by McVeigh (2004), who indicated that high impact OA journals are unevenly distributed among countries and research fields. Furthermore, the large presence of Central and South American journals (Giglia, 2010) could explain the large proportion of OA publications with low impact from countries such as Brazil, Spain, Chile or Argentina. Considering that there is an English-language positive bias and a Portuguese as well as Spanish-language negative bias in  the citation impact scores of OA journals (Ennas & Di Guardo, 2015), it is appropriate  to conclude  that "[t]he reasons behind [the] poor performance [of these countries] may be due to the national factor as well as to the large share of gold OA papers [compared to that of other] countries" (Torres-Salinas et al., 2016).

## 3.2. Characterizing OA journals by country and field

A next analysis focuses on the country of origin of gold OA journals. The purpose is to examine the characteristics of OA journals and illustrate how the relation between Gold OA impact, countries and fields is intimately related or closely linked with the presence of national OA journals from these countries and fields. The issue between countries' output and the country of origin of journals has been widely examined when studying the phenomenon of 'predatory journals' in OA publishing. Bohannon (2013) published a controversial experiment in Science magazine where he submitted hundreds of fake manuscripts to journals indexed in the *Directory for Open Access Journals* (DOAJ) and Bealls' list of predatory journals (Beall, 2012). His experiment criticized the lack of peer review of many of the journals to which he submitted his bogus papers which were, in many cases, accepted. This paper was seen by many as an attack to the Gold OA movement (Xia et al., 2015). Rapidly, studies were undertaken analyzing the extent to which 'predatory journals' were affecting the whole gold OA publishing enterprise. One of the most significant findings was that most of their action was geographically restricted to India and Nigeria (Shen & Björk, 2015; Xia et al., 2015).

Moreover, South America is the continent with the largest share of Gold OA publications (up to 74%) (Miguel, Chinchilla-Rodriguez, & de Moya-Anegón, 2011), but its presence in predatory journals barely represents 0.5% (Shen & Björk, 2015). These findings demonstrate that Gold OA publications are not necessarily of a lesser quality. Following what we observed in Figures 1 and 2, we could hypothesize that national differences in Gold OA impact are affected by disciplinary biases and the type of publisher of journals from these countries. Ennas & Di Guardo (2015) already point out that 'journals owned by UK and US publishers have a very strong and positive relation to the [positioning in the Scimago Journal Rank] ranking' and that 'journals adopting a business model requiring a form of payment to publish tend to become top rated more than others'. Similarly, Laakso & Björk (2012) highlight the diversity of types of journal publishers, geographical regions and scientific disciplines.

Table 2. Differences between Open Access journals and subscription-based journals. 2007-2016 period. Data retrieved from Web of Science

|  | Journals | Web of Science Docs | Avg Impact Factor | St dev Impact Factor | Avg CNCI | St dev CNCI |
|---|---|---|---|---|---|---|
| **Non-Open Access** | 10194 | 6554091 | 1.96 | 3.02 | 0.85 | 1.07 |
| **Open Access** | 1025 | 844036 | 1.94 | 2.22 | 0.62 | 0.59 |
| **Total general** | 11219 | 7398127 | 1.96 | 2.95 | 0.83 | 1.04 |

**Technical Note** Dataset: InCites Dataset; Schema: OECD; Document Type: Article; Time Period: 2007-2016
*InCites dataset updated May 13, 2017. Includes Web of Science™ content indexed through Mar 31, 2017.*

Table 2 compares the average Journal Impact Factor and Category Normalized Citation Impact (CNCI) of Gold OA journals, non-Gold OA journals and the total collection of journals. Interestingly, while there are no significant differences on the average Journal Impact Factor, Gold OA journals have on average a much lower Category Normalized Citation Impact than non-Gold OA journals (0.62 against 0.85).

Figures 4A and 4B compare the citation impact of Gold OA journals with that of other journals, broken down by discipline. They also indicate the absolute number of Gold OA journals in a discipline. Figure 4A shows results based on the journal impact factor, and Figure 4B on the category normalized citation impact. While Figure 4A shows that in Natural Sciences and in Humanities Gold OA journals have on average higher impact factor values than other journals have, and lower values in the other disciplines, Figure 4B reveals that CNCI values of Gold OA sources are below those of other journal in *all* disciplines, the difference being largest for Social Sciences and Humanities. Note that the lower number of OA journals in the humanities is largely due to the fact that Web of Science does not calculate the Impact Factor to journals indexed in the Arts & Humanities Citation Index.

Figure 4. Citation impact of Gold Open Access journals with that of other journals, broken down by discipline. AGR: Agricultural Science; ENG: Engineering; HUM: Humanities; MED: Medical Sciences; NAT: Natural Sciences; SOC: Social Sciences.

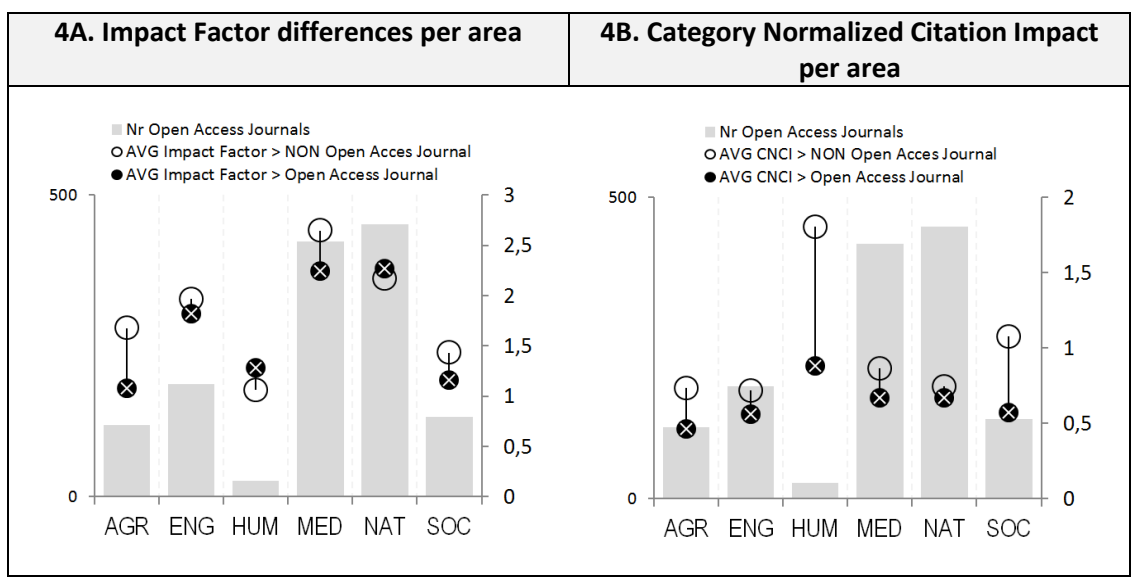

Figure 5 further analyses the category normalized citation impact (CNCI) of Gold OA journals in function of the country of the publisher of the journals. It reveals large differences in average CNCI between publishing countries. In the set of countries publishing more than 10 Gold OA journals, those published from The Netherlands, Germany, USA and England tend to have high CNCI values. These countries host large international publishing houses. Typical examples of countries with relatively low CNCI values and at least 5 Gold OA journals values are Colombia, Mexico, Serbia, India and Brazil. These evidences further strengthen the suggestion that the relation between impact and Gold OA could be more related to other factors such as type of publisher, country of the journal and the field of scope of the journal, rather than with the fact that they are OA journals. Indeed, this perception aligns well with the findings provided by Chavarro, Tang, & Rafols (2016) who showed that, in the case of Colombia, publishing in what they refer to as non-mainstream journals, is not only common, but that these journals are different in purpose to other journals, as they tend to publish research findings in subjects related to local knowledge or as a means to bridge between the international community and local communities.

Figure 5. Distribution and impact per country for OA journals indexed in the Web of Science

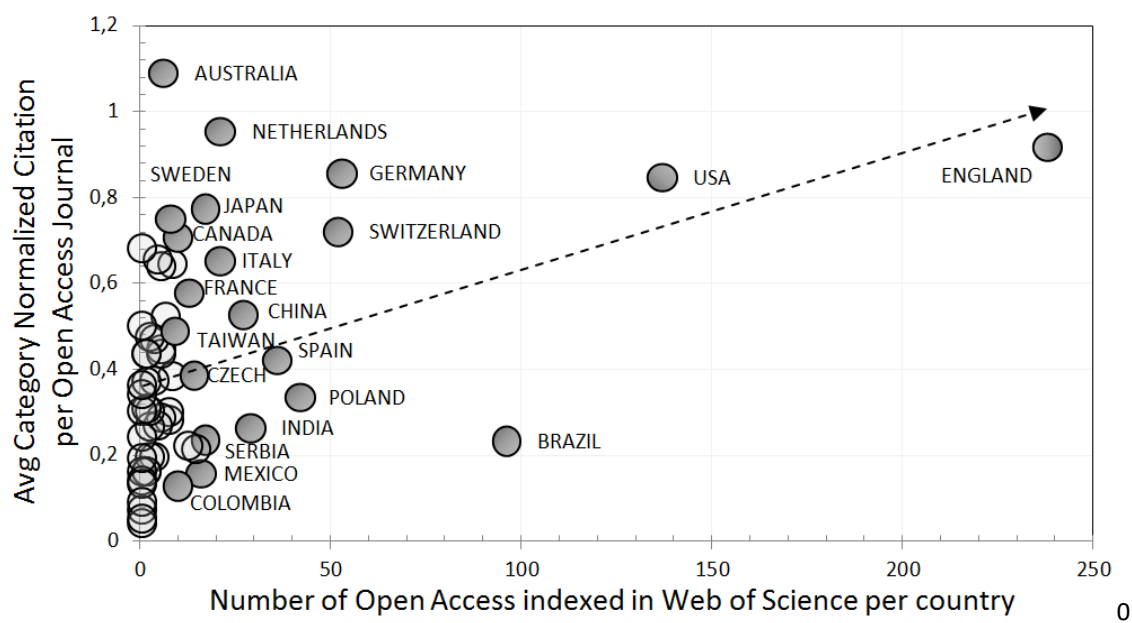

## 3.3. The effect of OA mega-journals in the publishing ecosystem

It is impossible to discuss Gold OA without taking some time to analyze the phenomenon of so-called OA Mega-journals (OAMJs). Defined as 'not only potentially disruptive in terms of altering the way research findings are assessed and communicated; [but] also disrupting academic culture itself' (Spezi et al., 2017), OAMJs have become, along with the raise of academic social media platforms (Björk, 2016), one of the major side-effects of Gold OA in the publishing industry. These journals have not only impacted into the scientific publishing culture due to the large number of papers they publish, but have also changed, or at least influenced, the ground rules of peer review by which a paper is considered to be worthy of publication.

Table 3. Articles published by Plos One by year, contribution to Gold OA overall publications and citation impact indicators. 2009-2016 period. Data retrieved from Web of Science

| Publication year | 2009 | 2010 | 2011 | 2012 | 2013 | 2014 | 2015 | 2016 | TREND |
|---|---|---|---|---|---|---|---|---|---|
| **Nr of documents by Plos One and contribution to the world** | | | | | | | | | |
| Web of Science documents by Plos One | 4403 | 6728 | 13780 | 23441 | 31492 | 30038 | 28114 | 22077 | 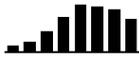 |
| % Contribution of Plos One to World´s OA totals | 7% | 9% | 15% | 20% | 22% | 18% | 15% | 12% | 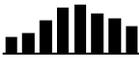 |
| **Plos One Impact Factor and Category Normalized Citation Impact trend** | | | | | | | | | |
| Journal Impact Factor | 4,35 | 4,41 | 4,09 | 3,73 | 3,53 | 3,23 | 3,06 | --- | 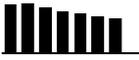 |
| Category Normalized Citation Impact | 1,64 | 1,51 | 1,37 | 1,24 | 1,13 | 1,04 | 0,83 | 0,66 | 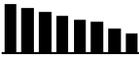 |

**Technical Note** Dataset: InCites Dataset; Schema: OECD; Document Type: Article; Time Period: 2007-2016
*InCites dataset updated May 13, 2017. Includes Web of Science™ content indexed through Mar 31, 2017.*

According to Wakeling et al. (2016), OAMJs are defined by two main characteristics: 1) they are broad in scope, covering many scientific areas, and 2) they set the peer review bar based on the technical soundness of manuscripts, not considering within their selection criteria the novelty of the publication or its significance and contribution to the field. This has been seen by many as a lowering of the publication standards, and partially explains the perception by many researchers that these journals are of a lesser quality. What is more, big publishers have now introduced their own mega-journals (e.g., Nature Communications, Scientific Reports or BMJ Open). This move has been seen by some as a way 'to tap into the stream of rejected manuscripts from their more selective top journals, in a system described as 'cascading reviews'' (Björk & Catani, 2016). Still, such malicious thinking does not seem supported by the data, which actually shows that, in terms of citations, both OAMJs and traditional journals show similar patterns (Björk & Catani, 2016).

Indeed, when examining the number of papers Plos One publishes every year, the share they represent from the Gold OA world output and its citation impact indicators, the numbers are quite revealing. Table 3 shows the publication trend of Plos One for the 2009-2016 period. Starting with 4,400 articles in the analyzed period, Plos One reached its highest peak of articles published by year in 2013, when it produced more than 30,000 publications. This number was relatively stable in the two subsequent years with a decrease to 22,000 publications in 2016. Plos One represented during the 2012-2014 period around 20% of all Gold OA publications, decreasing in the later years. This was due to the decrease of publications, but also to the increase of publications from Scientific Reports, which went from over 3,900 articles in 2014 to more than 10,700 articles in 2015 and up to 20,470 published papers in 2016.

A different issue is that related to the disruptive effect OAMJs have in scientific communities. According to Beall (2013), a strong opponent to Gold OA, '[t]hese journals, many of them now editorless, are losing the cohesion, soul, and community-binding roles that scholarly journals once had'. Traditionally, scientific journals have been considered as niches which tie and represent scientific communities, playing also a social role. OAMJs blur such communities as contents of different areas and communities are published in the same place. However, where some see a problem, others see an opportunity. MacCallum (2011), acknowledges such problem, but frames it from a different angle. OA means targeting not only scientists, but also policy makers, health managers, etc. OAMJs do not necessarily threaten the cohesion of scientific communities, but force publishers and information providers to rethink how to structure the increasing amount of literature produced by OAMJs not only 'to cater to different communities, but also to satisfy the needs of each individual and even enable them to generate new questions or discover novel avenues of research'. Probably a somewhat, opportunistic response to such issue, but one which reflects the disruptive effect OAMJs have had not only on the production of scientific literature, but also in its consumption.

# 4. CONCLUSIONS AND FUTURE PROSPECTS

Despite the large number of studies devoted to defining, characterizing, analyzing and discussing OA, its integration in the scholarly communication system is still a grand challenge which leaves room to further debate and discussion. More than 25 years of OA have gone by since the launch of ArXiv. Since then, topics under discussion have shifted many times. In the 1990s, studies tended to question copyright ownership (Bachrach et al., 1998) and explain the possibilities of making more accessible research findings (Ginsparg, 1997). The 2000s saw the expansion of OA, with many papers advocating for it (Crawford, 2002; Ginsparg, 2006; Swartz, 2008; Zuccala, 2009), explaining the different routes to OA (Harnad et al., 2008) and debating about whether OA gave greater visibility to research literature (Brody et al., 2006; Frandsen, 2009; Haque & Ginsparg, 2009; Moed, 2007), in many cases arguing in favor of a citation advantage (Craig et al., 2007; Davis & Fromerth, 2007; Davis, Lewenstein, Simon, Booth, & Connolly, 2008; Eysenbach, 2006; Gargouri et al., 2010).

In the last decade, new topics have been added to this on-going conversation. The settlement and growth of OA publishers, the emergence of OAMJs and of *predatory journals*, and the launch of academic social media platforms have led to more reflexive discussions as to the way OA is being integrated within the scientific communication system. Growing concerns as to how OA business models are affecting the quality of published research have been constantly present in the last few years (Beall, 2013; Björk & Solomon, 2012; Bohannon, 2013). This chapter intends to provide further insights as to the diversity of Gold OA and its citation impact in comparison with non-Gold OA. In this regard, the work made by Björk and colleagues (Björk et al., 2010; Björk & Solomon, 2012; Laakso et al., 2011; Laakso & Björk, 2012; Solomon et al., 2013) already provide great in-depth as to the heterogeneity of Gold OA journals and the many factors that could be affecting the negative view OA journals seem to provoke within a large sector of the scientific community (Agrawal, 2014).

Building from their findings, we can relate the following comments. Regarding the overall number of Gold OA publications and how much they represent from the overall number of publications, Björk et al. (2010) indicated that '8.5% of all scholarly journal volume for 2008 is available through some form of Gold OA'. The current study only focused on journal articles and it is based on data from Clarivate's InCites. We have obtained for 2008 a value of 4.37%, and for the year 2016 a value of 12.1%. This outcome illustrates that the overall share of Gold OA output is still increasing with the emergence of new players such as Scientific Reports or Nature Communications.

Björk & Solomon (2012) concluded in their 2012 study that 'OA journals indexed in Web of Science and/or Scopus are approaching the same scientific impact and quality as subscription journals, particularly in biomedicine and for journals funded by article processing charges'. In the current study, in which the Category Normalized Citation Impact (CNCI) of Gold OA articles is compared to that of all other articles, a moderate increase is found in the ratio of the impact of Gold OA and other types of articles. The question as to whether OA articles have a higher citation impact than non-OA papers has been addressed during the past 15 years in many studies analyzing the large multi-disciplinary citation indexes WoS and Scopus. Laakso et al., (2009), Bjork et al. (2010), and Bjork & Solomon (2012) present a review of these studies. To the best of the current authors' knowledge, the analysis presented in this chapter is one of the first to use the Category Normalized Citation Impact indicators at a large scale. The results show that Gold OA articles have in the 2012-2016 period on average a citation impact that is some 15% lower than the world average impact in the Gold OA articles' subject fields. It must be noted that this world average is based on all articles, both OA and non-OA.

The results presented in this chapter illustrate the heterogeneity in Gold OA publishing, and the skewness of the underlying publication output and citation impact distributions. A limited number of Gold OA journals accounts for a large percentage of the global Gold OA output, and the effects of these journals upon overall scores can be assumed to be substantial.

These outcomes illustrate how cautious one should be to draw generalized conclusions about Gold OA publishing. The results obtained in the current study confirm the conclusions by Björk & Solomon, (2012) that 'gold OA publishing is rapidly increasing its share of the overall volume of peer-reviewed journal publishing, and there is no reason for authors not to choose to publish in OA journals just because of the 'OA' label, as long as they carefully check the quality standards of the journal they consider'.

Table 4. Examples of countries representing three models of Gold OA publishing

| JOURNAL NAME | Publisher Country | Nr Articles | JIF | CNCI |
|---|---|---|---|---|
| MODEL 1 – UK – Publication in English language and high Impact Factor OA Journals | | | | |
| PLOS ONE | USA | 12948 | 3.057 | 1.14 |
| SCIENTIFIC REPORTS | ENGLAND | 3356 | 5.228 | 1.28 |
| NATURE COMMUNICATIONS | ENGLAND | 2131 | 11.329 | 3.48 |
| BMJ OPEN | ENGLAND | 2020 | 2.562 | 0.82 |
| JOURNAL OF HIGH ENERGY PHYSICS | ITALY | 1161 | 6.023 | 2.19 |
| BMC PUBLIC HEALTH | ENGLAND | 972 | 2.209 | 0.87 |
| NUCLEIC ACIDS RESEARCH | ENGLAND | 945 | 9.202 | 6.36 |
| FRONTIERS IN PSYCHOLOGY | SWITZER. | 863 | 2.463 | 1.2 |
| TRIALS | ENGLAND | 757 | 1.859 | 0.76 |
| ATMOSPHERIC CHEMISTRY AND PHYSICS | GERMANY | 686 | 5.114 | 2.24 |
| PLOS NEGLECTED TROPICAL DISEASES | USA | 679 | 3.948 | 1.8 |
| PLOS GENETICS | USA | 601 | 6.661 | 2.56 |
| NEW JOURNAL OF PHYSICS | ENGLAND | 595 | 3.57 | 1.34 |
| MALARIA JOURNAL | USA | 580 | 3.079 | 1.33 |
| PLOS PATHOGENS | USA | 543 | 7.003 | 2.58 |
| MODEL 2 – BRAZIL – Publication in national and low Impact Factor OA Journals | | | | |
| PLOS ONE | USA | 3961 | 3.057 | 0.86 |
| SEMINA-CIENCIAS AGRARIAS | BRAZIL | 1783 | 0.229 | 0.5 |
| CIENCIA RURAL | BRAZIL | 1746 | 0.376 | 0.22 |
| CIENCIA & SAUDE COLETIVA | BRAZIL | 1501 | 0.669 | 0.23 |
| ARQUIVO BRASILEIRO DE MEDICINA VETERINARIA E ZOOTECNIA | BRAZIL | 1192 | 0.21 | 0.16 |
| PESQUISA VETERINARIA BRASILEIRA | BRAZIL | 1057 | 0.335 | 0.23 |
| QUIMICA NOVA | BRAZIL | 1013 | 0.617 | 0.16 |
| CADERNOS DE SAUDE PUBLICA | BRAZIL | 950 | 0.92 | 0.3 |
| JOURNAL OF THE BRAZILIAN CHEMICAL SOCIETY | BRAZIL | 926 | 1.096 | 0.32 |
| REVISTA BRASILEIRA DE ENGENHARIA AGRICOLA E AMBIENTAL | BRAZIL | 889 | 0.478 | 0.25 |
| PESQUISA AGROPECUARIA BRASILEIRA | BRAZIL | 871 | 0.564 | 0.29 |
| REVISTA BRASILEIRA DE CIENCIA DO SOLO | BRAZIL | 748 | 0.611 | 0.36 |
| REVISTA DA ESCOLA DE ENFERMAGEM DA USP | BRAZIL | 748 | 0.415 | 0.15 |
| ANAIS DA ACADEMIA BRASILEIRA DE CIENCIAS | BRAZIL | 721 | 0.717 | 0.27 |
| BRAZILIAN JOURNAL OF BIOLOGY | BRAZIL | 686 | 0.559 | 0.18 |

| MODEL 3 – SPAIN – Publication in both, national and English language OA journals | | | | |
|---|---|---|---|---|
| PLOS ONE | USA | 5473 | 3.057 | 1.07 |
| SCIENTIFIC REPORTS | ENGLAND | 1273 | 5.228 | 1.37 |
| NUTRICION HOSPITALARIA | SPAIN | 892 | 1.497 | 0.3 |
| SENSORS | SWITZER. | 868 | 2.033 | 0.7 |
| JOURNAL OF HIGH ENERGY PHYSICS | ITALY | 613 | 6.023 | 2.45 |
| NATURE COMMUNICATIONS | ENGLAND | 608 | 11.329 | 3.47 |
| ANALES DE PSICOLOGIA | SPAIN | 490 | 0.574 | 0.36 |
| GACETA SANITARIA | SPAIN | 376 | 1.509 | 0.36 |
| PHYSICS LETTERS B | NETHER. | 362 | 4.787 | 3.48 |
| PSICOTHEMA | SPAIN | 340 | 1.245 | 0.62 |
| NEW JOURNAL OF PHYSICS | ENGLAND | 333 | 3.57 | 1.38 |
| SPANISH JOURNAL OF AGRICULTURAL RESEARCH | SPAIN | 322 | 0.76 | 0.37 |
| NUCLEIC ACIDS RESEARCH | ENGLAND | 317 | 9.202 | 1.9 |
| INFORMES DE LA CONSTRUCCION | SPAIN | 281 | 0.227 | 0.12 |
| BMC GENOMICS | ENGLAND | 276 | 3.867 | 1.28 |
| NEFROLOGIA | SPAIN | 276 | 1.207 | 0.35 |

The relationship between access modality and citation impact is complex, and does not allow for simple, general conclusions. One should be cautious with generalizing statements such as *OA journals have higher (or lower) impact than subscription-based serials*. Based on the findings presented here, we have observed three models of Gold OA production at the national level. These are shown in Table 4. The first model is that of countries like the USA, the United Kingdom, Germany or the Nordic countries, which publish in OA journals from big publishing firms with high Impact Factor, in many cases in OA mega-journals. The second model is exemplified in countries such as Brazil or India. These countries tend to have a large output in OA journals edited from their own countries. These journals tend to belong to specific fields (e.g., Agricultural Sciences in the case of Brazil), reinforcing the idea that they may serving as bridging between communities or focusing on topics of local or national interest (Chavarro et al., 2016). The last model is represented by countries like Spain or Poland which show a mixed combination of publishing in high impact OA mega-journals from big publishers as well as publishing in nationally-oriented OA journals from their own country.

These results illustrate the many factors that could affect the final citation impact of OA publishing and question statements against or in favor of OA publishing. Discussion related to Gold OA are not independent of other factors such as the disciplinary profile of countries' output, national characteristics or types of publishers.

## REFERENCES

Agrawal, A. A. (2014). Four more reasons to be skeptical of open-access publishing. *Trends in Plant Science*, *19*(3), 133. https://doi.org/10.1016/j.tplants.2014.01.005

Archambault, É., Amyot, D., Deschamps, P., Nicol, A., Provencher, F., Rebout, L., & Roberge, G. (2014). Proportion of open access papers published in peer-reviewed journals at the European and world levels—1996–2013. Retrieved from http://digitalcommons.unl.edu/scholcom/8/

Bachrach, S., Berry, R. S., Blume, M., Foerster, T. von, Fowler, A., Ginsparg, P., … Moffat, A. (1998). Who Should Own Scientific Papers? *Science*, *281*(5382), 1459–1460. https://doi.org/10.1126/science.281.5382.1459

Beall, J. (2012). Predatory publishers are corrupting open access. *Nature*, *489*(7415), 179.

Beall, J. (2013). The open-access movement is not really about open access. *TripleC: Communication, Capitalism & Critique. Open Access Journal for a Global Sustainable Information Society*, *11*(2), 589–597.

Björk, B.-C. (2014). Open access subject repositories: An overview. *Journal of the Association for Information Science and Technology*, 65(4), 698–706. https://doi.org/10.1002/asi.23021

Björk, B.-C. (2016). The open access movement at a crossroad: Are the big publishers and academic social media taking over? *Learned Publishing*, 29(2), 131–134. https://doi.org/10.1002/leap.1021

Björk, B.-C. (2017). Gold, green, and black open access. *Learned Publishing*, 30(2), 173–175. https://doi.org/10.1002/leap.1096

Björk, B.-C., & Catani, P. (2016). Peer review in megajournals compared with traditional scholarly journals: Does it make a difference? *Learned Publishing*, 29(1), 9–12. https://doi.org/10.1002/leap.1007

Björk, B.-C., & Solomon, D. (2012). Open access versus subscription journals: a comparison of scientific impact. *BMC Medicine*, 10, 73. https://doi.org/10.1186/1741-7015-10-73

Björk, B.-C., Welling, P., Laakso, M., Majlender, P., Hedlund, T., & Guðnason, G. (2010). Open Access to the Scientific Journal Literature: Situation 2009. *PLOS ONE*, 5(6), e11273. https://doi.org/10.1371/journal.pone.0011273

Bohannon, J. (2013). Who's Afraid of Peer Review? *Science*, 342(6154), 60–65. https://doi.org/10.1126/science.342.6154.60

Bohannon, J. (2016). Who's downloading pirated papers? Everyone. *Science*, 352(6285), 508–512. https://doi.org/10.1126/science.352.6285.508

Brody, T., Harnad, S., & Carr, L. (2006). Earlier Web usage statistics as predictors of later citation impact. *Journal of the American Society for Information Science and Technology*, 57(8), 1060–1072. https://doi.org/10.1002/asi.20373

Chavarro, D. A., Tang, P., & Rafols, I. (2016). *Why Researchers Publish in Non-Mainstream Journals: Training, Knowledge Bridging, and Gap Filling* (SSRN Scholarly Paper No. ID 2887274). Rochester, NY: Social Science Research Network. Retrieved from https://papers.ssrn.com/abstract=2887274

Chawla, D. S. (2017). Unpaywall finds free versions of paywalled papers. *Nature*. Retrieved from http://www.citeulike.org/group/10570/article/14339122

Cochran, A. (2014, August 8). Interview with Thomson Reuters: InCites Platform Offers New Analytics and Transparency. Retrieved 11 January 2018, from https://scholarlykitchen.sspnet.org/2014/08/08/qa-with-thomson-reuters-incites-platform-offers-new-analytics-and-transparency/

Craig, I. D., Plume, A. M., McVeigh, M. E., Pringle, J., & Amin, M. (2007). Do open access articles have greater citation impact?: A critical review of the literature. *Journal of Informetrics*, 1(3), 239–248. https://doi.org/10.1016/j.joi.2007.04.001

Crawford, W. (2002). Free electronic refereed journals: getting past the arc of enthusiasm. *Learned Publishing*, 15(2), 117–123. https://doi.org/10.1087/09531510252848881

Cullen, R., & Chawner, B. (2011). Institutional Repositories, Open Access, and Scholarly Communication: A Study of Conflicting Paradigms. *The Journal of Academic Librarianship*, 37(6), 460–470. https://doi.org/10.1016/j.acalib.2011.07.002

Davis, P. M., & Fromerth, M. J. (2007). Does the arXiv lead to higher citations and reduced publisher downloads for mathematics articles? *Scientometrics*, 71(2), 203–215. https://doi.org/10.1007/s11192-007-1661-8

Davis, P. M., Lewenstein, B. V., Simon, D. H., Booth, J. G., & Connolly, M. J. L. (2008). Open access publishing, article downloads, and citations: randomised controlled trial. *BMJ*, 337, a568. https://doi.org/10.1136/bmj.a568

Ennas, G., & Di Guardo, M. C. (2015). Features of top-rated gold open access journals: An analysis of the scopus database. *Journal of Informetrics*, 9(1), 79–89. https://doi.org/10.1016/j.joi.2014.11.007

Esposito, J. (2004). The devil you don't know: The unexpected future of Open Access publishing. *First Monday*, 9(8). https://doi.org/10.5210/fm.v9i8.1163

Eysenbach, G. (2006). Citation Advantage of Open Access Articles. *PLOS Biology*, 4(5), e157. https://doi.org/10.1371/journal.pbio.0040157

Frandsen, T. F. (2009). The effects of open access on un-published documents: A case study of economics working papers. *Journal of Informetrics*, 3(2), 124–133. https://doi.org/10.1016/j.joi.2008.12.002

Friend, F. J. (2003). Big Deal — good deal? Or is there a better deal? *Learned Publishing*, 16(2), 153–155. https://doi.org/10.1087/095315103321505656

Gargouri, Y., Hajjem, C., Larivière, V., Gingras, Y., Carr, L., Brody, T., & Harnad, S. (2010). Self-Selected or Mandated, Open Access Increases Citation Impact for Higher Quality Research. *PLOS ONE*, *5*(10), e13636. https://doi.org/10.1371/journal.pone.0013636

Gargouri, Y., Lariviere, V., Gingras, Y., Carr, L., & Harnad, S. (2012, September). *Green and Gold Open Access percentages and growth, by discipline*. Conference presented at the 17th International Conference on Science and Technology Indicators (STI). Retrieved from https://eprints.soton.ac.uk/340294/

Giglia, E. (2010). The impact factor of open access journals: data and trends. Retrieved from http://eprints.rclis.org/14666

Ginsparg, P. (1997). Winners and Losers in the Global Research Village. *The Serials Librarian*, *30*(3–4), 83–95. https://doi.org/10.1300/J123v30n03_13

Ginsparg, P. (2006). As We May Read. *Journal of Neuroscience*, *26*(38), 9606–9608. https://doi.org/10.1523/JNEUROSCI.3161-06.2006

Gumpenberger, C., Ovalle-Perandones, M.-A., & Gorraiz, J. (2013). On the impact of Gold Open Access journals. *Scientometrics*, *96*(1), 221–238. https://doi.org/10.1007/s11192-012-0902-7

Haque, A., & Ginsparg, P. (2009). Positional effects on citation and readership in arXiv. *Journal of the American Society for Information Science and Technology*, *60*(11), 2203–2218. https://doi.org/10.1002/asi.21166

Harnad, S., & Brody, T. (2004). Comparing the impact of open access (OA) vs. non-OA articles in the same journals. *D-Lib Magazine*, *10*(6). Retrieved from http://eprints.soton.ac.uk/260207

Harnad, S., Brody, T., Vallières, F., Carr, L., Hitchcock, S., Gingras, Y., … Hilf, E. R. (2008). The Access/Impact Problem and the Green and Gold Roads to Open Access: An Update. *Serials Review*, *34*(1), 36–40. https://doi.org/10.1080/00987913.2008.10765150

Jubb, M., Cook, J., Hulls, D., Jones, D., & Ware, M. (2011). Costs, risks and benefits in improving access to journal articles. *Learned Publishing*, *24*(4), 247–260. https://doi.org/10.1087/20110402

Kurtz, M., & Brody, T. (2006). The impact loss to authors and research. Retrieved from http://eprints.soton.ac.uk/40867/

Laakso, M., & Björk, B.-C. (2012). Anatomy of open access publishing: a study of longitudinal development and internal structure. *BMC Medicine*, *10*, 124. https://doi.org/10.1186/1741-7015-10-124

Laakso, M., Welling, P., Bukvova, H., Nyman, L., Björk, B.-C., & Hedlund, T. (2011). The Development of Open Access Journal Publishing from 1993 to 2009. *PLOS ONE*, *6*(6), e20961. https://doi.org/10.1371/journal.pone.0020961

MacCallum, C. J. (2011). Why ONE Is More Than 5. *PLOS Biology*, *9*(12), e1001235. https://doi.org/10.1371/journal.pbio.1001235

McVeigh, M. E. (2004). *Open access journals in the ISI citation databases: analysis of impact factors and citation patterns: a citation study from Thomson Scientific*. Thomson Scientific. Retrieved from http://www.academia.edu/download/39476484/openaccesscitations2.pdf

Meneghini, R., Mugnaini, R., & Packer, A. L. (2006). International versus national oriented Brazilian scientific journals. A scientometric analysis based on SciELO and JCR-ISI databases. *Scientometrics*, *69*(3), 529–538. https://doi.org/10.1007/s11192-006-0168-z

Miguel, S., Chinchilla-Rodriguez, Z., & de Moya-Anegón, F. (2011). Open access and Scopus: A new approach to scientific visibility from the standpoint of access. *Journal of the American Society for Information Science and Technology*, *62*(6), 1130–1145. https://doi.org/10.1002/asi.21532

Moed, H. F. (2007). The effect of "open access" on citation impact: An analysis of ArXiv's condensed matter section. *Journal of the American Society for Information Science and Technology*, *58*(13), 2047–2054. https://doi.org/10.1002/asi.20663

Piwowar, H., Priem, J., Larivière, V., Alperin, J. P., Matthias, L., Norlander, B., … Haustein, S. (2017). *The State of OA: A large-scale analysis of the prevalence and impact of Open Access articles* (No. e3119v1). PeerJ Preprints. https://doi.org/10.7287/peerj.preprints.3119v1

Shen, C., & Björk, B.-C. (2015). 'Predatory' open access: a longitudinal study of article volumes and market characteristics. *BMC Medicine*, *13*, 230. https://doi.org/10.1186/s12916-015-0469-2

Solomon, D. J., Laakso, M., & Björk, B.-C. (2013). A longitudinal comparison of citation rates and growth among open access journals. *Journal of Informetrics*, *7*(3), 642–650. https://doi.org/10.1016/j.joi.2013.03.008

Spezi, V., Wakeling, S., Pinfield, S., Creaser, C., Fry, J., & Willett, P. (2017). Open-access mega-journals: The future of scholarly communication or academic dumping ground? A review. *Journal of Documentation*, *73*(2), 263–283.

Suber, P. (2003). Removing the barriers to research: an introduction to open access for librarians. *College & Research Libraries News*, *64*. Retrieved from http://eprints.rclis.org/4616

Suber, P. (2005). Open access, impact, and demand: Why some authors self archive their articles. *BMJ: British Medical Journal*, *330*(7500), 1097.

Suber, P. (2008). Gratis and libre open access. Retrieved from https://dash.harvard.edu/handle/1/4322580

Swan, A. (2010). The Open Access citation advantage: Studies and results to date. Retrieved from http://eprints.soton.ac.uk/268516

Swartz, A. (2008). Guerilla open access manifesto. *Online-Ressource, URL Http://Www. Openeverything. Eu/Guerilla-Open-Access-Manifest*. Retrieved from http://wavelets.ens.fr/BOYCOTT_ELSEVIER/DECLARATIONS/DECLARATIONS/2008_07_01_Aaron_Swartz _Open_Access_Manifesto.pdf

Tennant, J. P., Waldner, F., Jacques, D. C., Masuzzo, P., Collister, L. B., & Hartgerink, C. H. J. (2016). The academic, economic and societal impacts of Open Access: an evidence-based review. *F1000Research*, *5*, 632. https://doi.org/10.12688/f1000research.8460.3

Torres-Salinas, D., Robinson-Garcia, N., & Aguillo, I. F. (2016). Bibliometric and Benchmark Analysis of Gold Open Access in Spain: Big Outpu... *Profesional de La Informacion*, *25*(1), 17–24. https://doi.org/10.3145/epi.2016.ene.03

van Leeuwen, T. N., Tatum, C., & Wouters, P. (2015). Open Access Publishing and Citation Impact-An International Study. In *ISSI*. Retrieved from https://pdfs.semanticscholar.org/55f9/1f879fe7867da70b782914ac78d6fc7c6984.pdf

Van Noorden, R. (2013a). Half of 2011 papers now free to read. *Nature News*, *500*(7463), 386. https://doi.org/10.1038/500386a

Van Noorden, R. (2013b). Open access: The true cost of science publishing. *Nature News*, *495*(7442), 426. https://doi.org/10.1038/495426a

Wakeling, S., Willett, P., Creaser, C., Fry, J., Pinfield, S., & Spezi, V. (2016). Open-Access Mega-Journals: A Bibliometric Profile. *PLOS ONE*, *11*(11), e0165359. https://doi.org/10.1371/journal.pone.0165359

Whitfield, J. (2012). Elsevier boycott gathers pace. *Nature*, *9*. https://doi.org/10.1038/nature.2012.10010

Xia, J., Harmon, J. L., Connolly, K. G., Donnelly, R. M., Anderson, M. R., & Howard, H. A. (2015). Who publishes in "predatory" journals? *Journal of the Association for Information Science and Technology*, *66*(7), 1406–1417. https://doi.org/10.1002/asi.23265

Zuccala, A. (2009). The lay person and Open Access. *Annual Review of Information Science and Technology*, *43*(1), 1–62. https://doi.org/10.1002/aris.2009.1440430115